\documentclass[epj]{svjour}
\usepackage{graphics}
\begin{document}
\title{Determining the inelastic proton-proton cross section at the Large Hadron Collider using minimum bias events.}
\titlerunning{Determining the inelastic proton-proton cross section at the LHC.}
\author{I. Dawson, K. Prokofiev
}                     
%
%
\institute{Department of Physics and Astronomy, University of Sheffield, Sheffield, S3 7RH, UK}
%
%
\abstract{
Described in this paper is a new method for determining the non-diffractive part of the 
inelastic proton-proton cross section, at the LHC centre of mass energy of $14$~TeV. 
The method is based on counting the number of inelastic proton-proton interactions in 
the collision regions. According to preliminary investigation, this measurement will be best suited 
for the initial low luminosity phase of the LHC. 
The dominant uncertainty on this measurement comes from knowledge of the proton-proton luminosity. 
%
} 
\maketitle
\section{Introduction}\label{Sec:1}
The Large Hadron Collider (LHC) has been designed to collide protons at a centre of mass energy 
$\sqrt{s}=14$\,TeV, providing experiments with a TeV-scale reach for new physics. 
Also possible at the LHC are measurements of soft interaction physics processes, 
using so called minimum bias data \cite{OurPaper}. 
The study of soft interactions in proton-proton interactions at $\sqrt{s}=14$\,TeV, 
and comparison with results obtained at lower centre of mass energies, should improve our understanding of 
the structure of the proton in the non-perturbative QCD regime.

The proton-proton ($pp$) interaction rate at the LHC is given by 
\begin{equation}
 N=\sigma_{pp}\cdot L,
 \label{eq:num_col}
\end{equation}
 where $\sigma_{pp}$ is the total $pp$ 
cross section and $L$ is the luminosity.
The cross section has contributions from elastic, diffractive 
and non-diffractive inelastic processes \cite{GoodBook}, 
so that $\sigma_{pp} = \sigma_{el} + \sigma_{di} + \sigma_{nd}$. 
The diffractive cross section $\sigma_{di}$ can be subdivided further, with the dominant contributions 
coming from single and double diffractive dissociation. 

The CDF experiment measured the total and elastic antiproton-proton cross sections at $\sqrt{s}=1.8$\,TeV
to be $80.03 \pm 2.24$\,mb and $19.70 \pm 0.85$\,mb respectively \cite{CDF1,CDF2}. 
Also at $\sqrt{s}=1.8$\,TeV, the E710 experiment measured the total cross section to be 
$72.8 \pm 3.1$\,mb \cite{E710}. 
These measurements confirmed the rise in total cross section observed at lower centre 
of mass energies. However the difference in the CDF and E710 results means that 
the form of the rise in total cross section is ambiguous at the TeV scale. 
Cross section measurements at the LHC energy $\sqrt{s}=14$\,TeV should therefore provide an important 
insight into the nature of this rise.

Described in this paper is a new method for measuring the non-diffractive part of the inelastic 
$pp$ cross section, using minimum bias events. A description of minimum bias events and their selection 
is given in Sec.~\ref{Sec:2}.
The method is based on determining the mean number of inelastic $pp$ collisions in a bunch crossing. 
In the interaction regions of the LHC experiments ATLAS and CMS, nearly every bunch 
crossing will contain at least one inelastic $pp$ collision. The subject of collision multiplicity is 
discussed in Sec.~\ref{Sec:3}. 
If individual collisions can be identified and counted, then the mean can be determined.
We use the example of the ATLAS experiment to show in Secs.~\ref{Sec:4}-\ref{Sec:6} that the 
identification of each collision should be possible using charged particles produced within the 
acceptance of the Inner Detector. 
A discussion of uncertainties is given in Sec.~\ref{Sec:7}, and in Sec.~\ref{Sec:8} we present our conclusions.

\section{Minimum bias events}\label{Sec:2}
In hadron collider experiments, inelastic events are selected using a minimum bias (MB) trigger. 
Typically the MB trigger accepts non-diffractive and double diffractive inelastic events, but rejects 
single diffractive events. The MB events are therefore commonly identified with non-single 
diffractive inelastic (NSD) events.

At the LHC, the inelastic collision can also be collected by triggering on 
randomly selected bunch crossings.
These data will contain both non-diffractive and diffractive events, which can be analysed offline. 
Single diffractive events can be removed to create NSD data sets similar to the MB trigger data. 
An advantage of the random trigger is that no trigger bias is introduced in the event selection. 
In this paper, events selected with either the MB trigger or the random trigger are referred to as MB events. 
It is assumed that non-diffractive inelastic events can be extracted from MB events, although 
the double diffractive component in the NSD data will require further consideration.

\section{Collision multiplicities}\label{Sec:3}
The design luminosity at the LHC is $10^{34}$\,cm$^{-2}$\,s$^{-1}$, corresponding to an average proton 
luminosity per bunch crossing of approximately 3$\times10^{26}$\,cm$^{-2}$ \cite{LHC}.  
The number of interactions per bunch crossing will vary according to Poisson statistics,
with a mean value determined by Eq.~\ref{eq:num_col}.
For a cross section $\sigma_{nd} = 60$~mb, this would give a mean of $18$ inelastic non-diffractive 
events per bunch crossing. 
During the early stages of the LHC, the plan is to run at the lower luminosity of 
$2\times10^{33}\,{\rm cm}^{-2}{\rm s}^{-1}$. Assuming the same number of proton bunches, this would give 
an average of $3.6$ interactions per bunch crossing in the case of $\sigma_{nd} = 60$~mb. We refer to this initial 
luminosity phase as low luminosity, and high luminosity for the case of $L$\,$=$\,$10^{34}$\,cm$^{-2}$\,s$^{-1}$.

\begin{figure}
 \resizebox{0.5\textwidth}{!}{\includegraphics{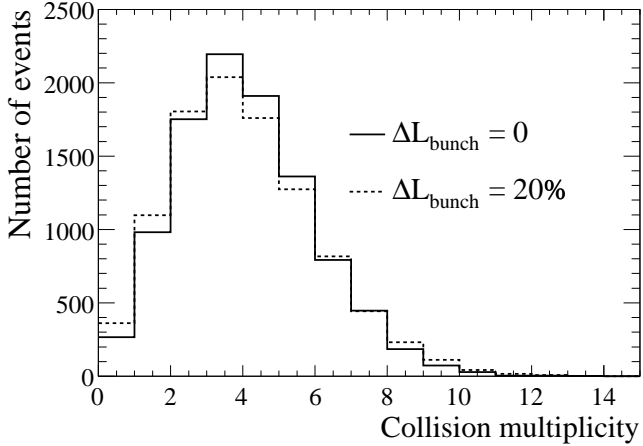}}
  \caption{Number of inelastic non-diffractive $pp$ interactions per bunch 
           crossing at low luminosity, assuming $\sigma_{nd} = 60$~mb. The mean multiplicity is 3.6.}
   \label{fig:2}
 \end{figure}

Shown in the Fig.~\ref{fig:2} is a Poisson distribution representing the number of
interactions per bunch crossing for the low luminosity case (solid line).  
The dashed line represents the distribution obtained from varying the mean of the Poisson according to a Gaussian 
distribution. This latter case represents a 20\% uncertainty in luminosity for different proton bunches. 
It can be noted that while the variance of the multiplicity distribution changes, the mean does not.

\section{Particle multiplicity}\label{Sec:4}
As discussed above, the proposed method is based on determining the mean
number of $pp$ collisions in a bunch crossing. This requires identifying and counting the number 
of primary vertices of $pp$ interactions with high efficiency.
A successful reconstruction of primary vertices requires that in each collision at least several 
charged particles are produced within the acceptance of a detector and that the reconstruction of
their trajectories is efficient.

In the paper of \cite{OurPaper}, the predictions of two Monte Carlo event generators PYTHIA \cite{PYTHIA} 
and PHOJET \cite{PHOJET} were compared with minimum bias and underlying event data for a wide range of centre of 
mass energies. Both of these generators have successful models for describing both the soft and hard scattering 
components in inelastic $pp$ interactions. A new set of tunings were developed and used to extrapolate the Monte 
Carlo predictions to LHC energies. 

\begin{figure}
\resizebox{0.5\textwidth}{!}{
  \includegraphics{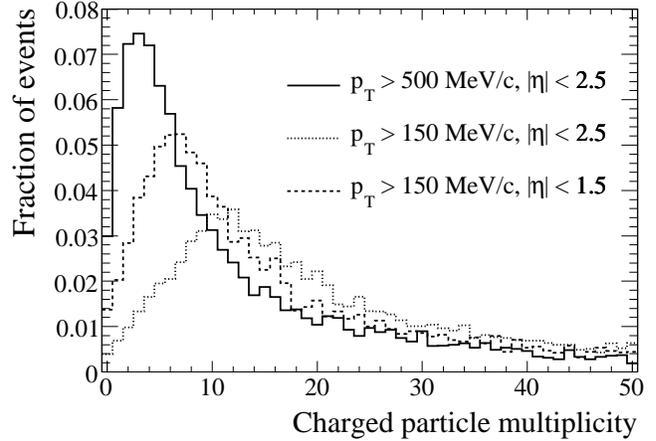}
}
\caption{Number of charged particles produced in inelastic 
non-diffractive $pp$ interactions at $\sqrt{s}= 14$~TeV, 
normalised to the number of events, as predicted by PYTHIA~6.214.}
\label{fig:1}
\end{figure}

Shown in Fig.~\ref{fig:1} are the distributions of charged particle multiplicity for inelastic non-diffractive 
events, as predicted by PYTHIA~6.214. The distributions are presented for three different combinations of cuts 
on particle transverse momentum $p_T$ and pseudorapidity $\eta$\footnote{The angular coverage of the ATLAS and 
CMS Inner Detectors corresponds to a pseudorapidity range $|\eta|<2.5$, where $\eta = -\ln(\tan(\theta/2))$ and 
$\theta$ is the polar angle to the beam axis.}. These combinations correspond to different regions of phase space 
where track reconstructing should be possible with the tracking detector of ATLAS.
  
It can be noted that within the angular acceptance 
of $|\eta|<2.5$, the average numbers of charged particles per event for transverse momenta 
$p_T > 500$~MeV/c and $p_T > 150$~MeV/c are $15.4$ and $36.4$ respectively. 
It is also important that the probability of finding no particles per collision is small: 
$\sim 3$\% for the case of $p_T  >500$~MeV/c, and only $\sim 0.4$\% for the case of $p_T >150$~MeV/c.
These two facts give the motivation to reconstruct tracks starting
from the lowest $p_T$ possible. In addition, the fact that at least one charged particle 
will be produced in nearly every $pp$ collision makes efficient primary vertex finding possible.

Also shown in Fig.~\ref{fig:1} is the multiplicity distribution for 
charged particles with $p_T > 150$~MeV/c produced in the 
reduced pseudorapidity range $|\eta|<1.5$ (usually referred to as 
the central tracking region). The
number of particles produced in $pp$ event is $21.8$. It is 
expected that trajectories of charged particles reconstructed 
in the central region of the ATLAS tracker will have a better 
position resolutions \cite{ATLAS-Phys}, allowing for more accurate vertex counting.
This point is discussed further in the following sections.


\section{Track reconstruction}\label{Sec:5}
For the purpose of vertex counting, it is important that the trajectories of charged particles 
are reconstructed efficiently and fake rates are kept low. In general, reconstruction 
inefficiencies are related to the performance of detector and track reconstruction algorithms.
In the case of charged pions, which dominate the spectrum of charged particles in inelastic 
$pp$ collisions, there are also losses of efficiency due to hadronic interactions 
in the detector material and decays. According to ATLAS performance simulations \cite{ATLAS-Phys}, reconstruction 
efficiencies of approximately $80$\% are obtained for charged pions with transverse momenta $p_T>500$~MeV/c.

The resolution on the reconstructed track parameters is also crucial for vertex counting. 
Of particular importance is resolution on the $z_0$ impact parameter\footnote{The $z_0$ 
impact parameter is the longitudinal (along the beam axis) distance between the point of the 
closest approach of a reconstructed trajectory of a charged particle and a reference point 
with respect to which the parameters of this trajectory are calculated.}. In order to distinguish 
different $pp$ collisions within the same bunch crossing, we need the resolution on $z_0$ to be smaller
than the typical distance between adjacent interactions.
For charged tracks with $p_{T} = 500$~MeV/c produced in the central tracking region of the ATLAS 
Inner Detector, the resolution on the $z_0$ impact parameters is found  to be 
$\sigma(z_{0})\approx 200\,\mu {\rm m}$ \cite{ATLAS-Phys}. However, the resolution degades rapidly 
for $|\eta| > 1.5$, and $\sigma(z_{0})$ reaches a few mm in the most forward regions.

As discussed in Sec.~\ref{Sec:4}, there is motivation to reconstruct charged tracks with transverse 
momenta below $p_{T} = 500$~MeV/c. The trajectory of a charged particle 
with $p_{T} = 150$~MeV/c, travelling in a $2$ tesla magnetic field, has a transverse radius of 
curvature of approximately $25$~cm. Ignoring decays and inelastic hadron interactions in the 
detector material, this particle will reach  the outer layers of the ATLAS silicon tracker, 
providing up to $7$ measurements along its trajectory \cite{ATLAS-Phys}. Therefore it should be 
possible to reconstruct particle trajectories with transverse momenta significantly below $p_{T} = 500$~MeV/c. 


\section{Primary vertex counting}\label{Sec:6}
The number of inelastic $pp$ interactions in a bunch crossing will vary according to Poisson 
statistics, as discussed in Sec.~\ref{Sec:2}. The space distribution of the interactions produced 
in a bunch crossing will depend on the proton bunch profile. In the direction of the beam 
(defined as the $z$-axis) the collisions in bunch crossing will be distributed according to a Gaussian 
with standard deviation $\sigma_{z}\approx 5.6$~cm. The transverse dimensions of a bunch crossing 
region $\sigma_{x}$ and $\sigma_{y}$ are much smaller, $\sim 15\,\mu{\rm m}$. 
If the displacement in $z$ between two adjacent $pp$ interactions in a bunch crossing is greater than the 
resolutions on the $z_0$ impact parameter of the reconstructed tracks, then associating tracks to vertices 
should be possible.

\begin{figure}
\resizebox{0.5\textwidth}{!}{%
  \includegraphics{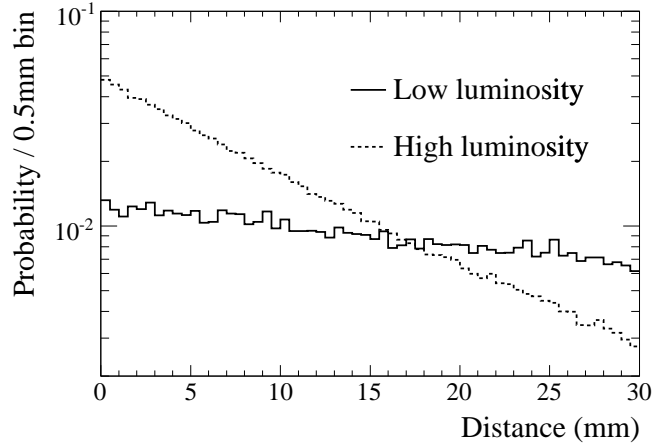}
}
\caption{Distance between pairs of adjacent $pp$ interactions, for bunch crossings with two or more interactions.}
\label{fig:3}
\end{figure}

Shown in the Fig.~\ref{fig:3} are the distances between pairs of adjacent $pp$ interactions 
in an LHC bunch crossing, for $\sigma_{z}= 5.6$~cm, for the high and low luminosity cases.
In the low luminosity case, the probability of finding two interactions closer than $0.5$~mm ($5$~mm)
to each other is $\sim 1$\% ($\sim 10$\%). 
As discussed in Sec.~\ref{Sec:5}, in the central region of the ATLAS Inner Detector the resolution 
on the $z_0$ impact parameter for tracks with $p_{T} = 500$~MeV/c is about $200\,\mu{\rm m}$.
A correct association of tracks to their production vertices should thus be possible.
In the high luminosity mode, the reconstruction of the number of interaction vertices should also be 
possible in ATLAS, given that the resolution on the $z_{0}$ impact parameter is not expected to degrade.

As mentioned above, $\sigma(z_{0})$ increases in the most forward regions 
up to a few mm. It is expected that the efficiency of the vertex finding will degrade due to
the charged tracks produced in this region. The optimal geometrical cuts required for the effective 
vertex finding should be established through a dedicated analysis.
It should be noted however, that in the low luminosity regime, the degradation in question will be 
predominantly small. A careful inspection of the Fig.~\ref{fig:3} shows that only $\sim$$10$\% of pairs 
of adjacent $pp$ interactions will have distances less than $5$~mm.

\section{Uncertainties}\label{Sec:7}
The main uncertainties associated with using the vertex counting method to determine 
the inelastic $pp$ cross section are the following:
\begin{enumerate}
\item {\it The $pp$ luminosity.} 
At the LHC the expected precision for measuring the 
average luminosity is $5 - 10\%$~\cite{ATLAS-Lumi}. This is likely to dominate the inelastic $pp$ cross section 
measurement uncertainty. 

The uncertainty associated with fluctuations on the bunch crossing luminosity is expected to be small. 
If the fluctuations are distributed symmetrically around a mean, then the mean value 
of the event multiplicity distribution does not change, as discussed in Sec.~\ref{Sec:3}.
A Gaussian distribution is anticipated due to the many proton bunches at the LHC. The variance is not 
expected to exceed $20$\%. Indeed, measuring the deviation of the event multiplicity variance from 
that of a Poisson distribution will give a measure of the bunch-bunch luminosity fluctuations.
\item {\it Diffractive events.} 
The predictions of PHOJET for single and double diffractive cross sections in $pp$ scattering at the 
LHC are 11\,mb and 4.1\,mb respectively. These values can be compared to the prediction for the non-diffractive 
inelastic cross section of 69.5\,mb. The event topology of diffractive events is distinctive because of 
so called rapidity gaps, where particles are produced in the very forward regions with little event 
activity in the central rapidity region. In contrast, particles are produced predominantly in the 
central rapidity region in non-diffractive inelastic events.

Determining the fraction of diffractive events in the minimum bias data sample, and their rejection, 
will require a dedicated analysis. The contamination in the final data sample is expected to be 
small, along with the corresponding uncertainty on the cross section measurement. We note that at the UA1 experiment 
at $\sqrt{s}=0.9$\,TeV the background contribution from single diffractive events in their minimum bias data 
was estimated to be less than 2\% \cite{UA1}. 

\item {\it Beam gas interactions.} The contribution of beam-gas interactions in MB data samples 
will have to be studied. This background is usually easy to discriminate against due to the 
asymmetry in  pseudorapidity. For example, the CDF experiment estimated the contamination of misidentified 
beam-gas interactions in their final data sample to be less than 0.5\% at $\sqrt{s}=1.8$\,TeV and less 
than 2.5\% at $\sqrt{s}=600$\,GeV.
\item {\it Detector acceptance.} Not all inelastic events result in particles being produced 
within the detector acceptance. 
However, as discussed in Sec.~\ref{Sec:4}, less than 5\% of events have no particle 
produced within the detector acceptance for $p_T > 500$~MeV/c. For the lower transverse momenta,
the fraction of these events will be even smaller. It can be therefore concluded that the corrections
associated with detector acceptance will be small and so will be the corresponding 
uncertainty on the cross section measurement.
\item {\it The vertex counting efficiency.} The uncertainty related to the efficiency of vertex 
counting depends strongly on the resolutions on the $z_0$ impact parameter of the reconstructed tracks.
In order to estimate this contribution, a dedicated analysis is required. 
However, as mentioned in the Sec.~\ref{Sec:6}, only $\sim 10$\% of $pp$ interactions
will have distances within $5$~mm of each other along the $z$-axis. This distance is 
much bigger than the typical $z_0$ resolutions of tracks, discussed in Sec.~\ref{Sec:4}.
The associated uncertainty is therefore not expected to be more than a few percent.
\end{enumerate}

\section{Conclusions}\label{Sec:8}
In this paper we have described a new method for measuring the inelastic non-diffractive cross section, 
which can be applied at LHC experiments. We have argued that the accuracy of the measurement will be at the level 
of $5 - 10\%$ and that the dominant uncertainty will come from knowledge of the luminosity. 

We have shown that the ATLAS experiment provides an excellent laboratory for testing this new approach. 
We have also suggested that the reconstruction of particle trajectories to lower transverse momenta 
should be beneficial for finding the primary vertices 
associated with the inelastic $pp$ collisions. We therefore recommend the development of track reconstruction 
algorithms 
into this low-$p_T$ regime. 
Similar arguments apply to the CMS experiment, although its prospects for low-$p_T$ track measurement are 
more challenging due to the higher magnetic field in the CMS Inner Detector. 

We note that measurements of the total and elastic cross sections are being planned by the TOTEM
experiment at the LHC using a luminosity independent method based on the Optical Theorem \cite{TOTEM}. 
These measurements require dedicated instrumentation in the very forward regions. Diffractive physics measurements 
will also feature strongly. The inelastic non-diffractive cross section can be obtained by subtracing the elastic and 
diffractive components from the total cross section. 
This measurement should complement our more direct measurement. 

Finally we would like to point out that, while we have proposed the vertex counting method for making a measurement 
of the inelastic non-diffractive cross section, this method can be used equally for luminosity 
measurement. Once we have some knowledge of $\sigma_{nd}$ at $\sqrt{s}=14$\,TeV, either from ATLAS or from TOTEM, then 
the efficient reconstruction of the number of minimum bias events should allow measurements of the LHC luminosity.


\begin{thebibliography}{}
\bibitem{OurPaper}
A. Moraes {\it et al}., Eur. Phys. J. {\bf C} (2007)
\bibitem{GoodBook}
P. Collins and A. Martin, Rep. Prog. Phys. {\bf 45}, (1982) 335
\bibitem{CDF1}
F. Abe {\it et al}., Phys. Rev. D. {\bf 50}, (1994) 5550
\bibitem{CDF2}
F. Abe {\it et al}., Phys. Rev. D. {\bf 50}, (1994) 5518
\bibitem{E710}
N. M. Amos {\it et al}., Phys. Rev. Lett. {\bf 68}, (1992) 2433
\bibitem{LHC}
The Large Hadron Collider, CERN/AC/95-05, (1995)
\bibitem{PYTHIA} 
T. Sj\"ostrand {\it et al.}, Comput. Phys. Comm. {\bf 135}, (2001) 238  
\bibitem{PHOJET} 
R. Engel, Z. Phys. C. {\bf 66}, (1995) 203 
\bibitem{ATLAS-Phys}
ATLAS Detector and Physics Performance TDR, Vol. 1, Chapter 3, CERN/LHCC/99-14 (1999)
\bibitem{ATLAS-Lumi}
ATLAS Detector and Physics Performance TDR, Vol. 1, Chapter 13, CERN/LHCC/99-14 (1999)
\bibitem{UA1}
C. Albajar {\it et al}., Nucl. Phys. B. {\bf 336}, (1990) 261
\bibitem{TOTEM}
TOTEM TDR, CERN/LHCC/2004-002 (2004)
\end{thebibliography}
\end{document}